\documentclass[12pt,a4paper]{article}
\usepackage{amsfonts,latexsym}
\usepackage{graphicx,color}
\usepackage{dcolumn}
\usepackage{graphicx}
\usepackage{amsmath}
\usepackage{amsfonts}
\usepackage{amssymb}
\usepackage{psfrag}
\usepackage{wrapfig}
\usepackage{subfigure}
\usepackage{makeidx}

\renewcommand{\thefootnote}{\fnsymbol{footnote}}

\begin{document}

\vspace{12mm}

\begin{center}
{{{\Large {\bf Scalarized black holes in the Einstein-Maxwell-scalar theory with a quasi-topological term }}}}\\[10mm]

{Yun Soo Myung$^a$\footnote{e-mail address: ysmyung@inje.ac.kr} and De-Cheng Zou$^{a,b}$\footnote{e-mail address: dczou@yzu.edu.cn}}\\[8mm]

{${}^a$Institute of Basic Sciences and Department  of Computer Simulation, Inje University Gimhae 50834, Korea\\[0pt] }

{${}^b$Center for Gravitation and Cosmology and College of Physical Science and Technology, Yangzhou University, Yangzhou 225009, China\\[0pt]}
\end{center}
\vspace{2mm}

\begin{abstract}
We investigate the Einstein-Maxwell-scalar theory with a quasi-topological term.
Considering exponential couplings to Maxwell term, quasi-topological term, and both terms,  we obtain three sets of  infinite scalarized charged  black holes
by taking into account tachyonic instability of dyonic Reissner-Nordstr\"{o}m black hole.
Each set of infinite scalarized  charged   black holes is classified   by the number of $n=0,1,2,\cdots$, where $n=0$ is called the fundamental black hole and $n=1,2,\cdots$ denote the $n$-excited black holes. All $n=0$ black holes are stable against the radial perturbation, while all $n=1,2$ black holes are unstable.
\end{abstract}
\vspace{5mm}

\vspace{1.5cm}

\hspace{11.5cm}
\newpage
\renewcommand{\thefootnote}{\arabic{footnote}}
\setcounter{footnote}{0}

\section{Introduction}
There exist three types of bosonic fields which were used in the physical description of nature as spin-0,-1, and -2 fields.
In the context of strong  gravity (black hole spacetimes), the interplay between the gravitational, electromagnetic, and scalar fields  has a long history.
No-hair theorem states that a black hole is completely described by the mass, electric charge, and angular momentum~\cite{Ruffini:1971bza}.
In this connection, the gravitational and electromagnetic  fields
satisfy the Gauss-law outside the horizon. If a scalar field is minimally coupled to  these fields, it could not exist as an equilibrium configuration
around the black hole, implying no scalar hair~\cite{Herdeiro:2015waa}. Surely, a minimally coupled scalar does not satisfy the Gauss-law.

However, if one introduces  non-minimal couplings,   black holes with scalar hair might be  allowed.
Considering the Einstein-conformally coupled scalar theory,
one has a secondary scalar-hair around the BBMB black hole  although the scalar hair blows up on the horizon~\cite{Bocharova:1970skc,Bekenstein:1974sf}.
Also, the dilaton has a specific non-minimal coupling to the Maxwell term, leading to  charged black holes with scalar hair~\cite{Gibbons:1987ps,Garfinkle:1990qj,Gibbons:1982ih}.

Recently, newly black holes with scalar hair were found  from the Einstein-Gauss-Bonnet-scalar (EGBS) theory~\cite{Doneva:2017bvd,Silva:2017uqg,Antoniou:2017acq} and Einstein-Maxwell-scalar (EMS) theory~\cite{Herdeiro:2018wub}
by introducing  the non-minimal couplings  to the Gauss-Bonnet term ($f(\phi){\cal G}$) and Maxwell term ($f(\phi) F^2$)~\cite{Myung:2018vug,Myung:2018jvi,Myung:2019oua}.
It is worth mentioning that the onset of scalarization  is surely captured by linearized theory and thus, it is  implemented by the tachyonic instability of black holes without scalar hair. The end point of the instability depends on  nonlinear interactions of $f(\phi)=\alpha \phi^2$ and $e^{\alpha \phi^2}$ which eventually quench the linear instability.
Here, we would like to  stress that  it is dubbed the spontaneous scalarization.

In this work, we will introduce  a generalization of the EMS theory to investigate the spontaneous scalarization furthermore. It describes a real scalar field minimally coupled to the Einstein gravity while it couples  non-minimally to the Maxwell  term $F^2$, the quasi-topological term $((F^2)^2-2F^{(4)})$~\cite{Liu:2019rib}, and both terms.
It is worth reminding  that  the quasi-topological term is regarded as  a square of the topological term  ($F\tilde{F}$)~\cite{Fernandes:2019kmh}.
We note that the topological term does not alter the black hole solution, while the quasi-topological term modifies the  black hole solution when considering
the dyonic Reissner-Norstr\"{o}m (RN) black hole~\cite{Astefanesei:2019pfq}.

Introducing exponential coupling $e^{\alpha \phi^2}$ to Maxwell term, quasi-topological term, and both terms,  we obtain three sets of  $n=0,1,2,\cdots$ scalarized charged  black holes
by taking into account tachyonic instability of dyonic Reissner-Nordstr\"{o}m black hole.
The stability of all scalarized charged black holes with respect to radial perturbations will be reached by examining
the qualitative behavior of the potential  as well as by obtaining  exponentially growing (unstable) modes for $s$-mode scalar  perturbation.
We find that all $n=0$ black holes are stable, whereas all $n=1,2$ black holes are unstable.

\section{Scalar coupling to Maxwell term}
\subsection{Dyonic RN black hole }

Let us introduce  the action for the EMS theory with quasi-topological electromagnetism
\begin{equation}
S_{\rm EMSt}=\frac{1}{16 \pi}\int d^4 x\sqrt{-g}\Big[ R-2\partial_\mu \phi \partial^\mu \phi-e^{\alpha \phi^2} F^2-\Big((F^2)^2-2F^{(4)}\Big)\Big],\label{Act1}
\end{equation}
where $\alpha$ is a scalar coupling parameter to the Maxwell term, $F^2=F_{\mu\nu} F^{\mu\nu}$,  and $F^{(4)}=F^\mu_\nu F^\nu_\rho F^\rho_\sigma F^\sigma_\mu$.
Here, the last term denotes the quasi-topological electromagnetism which is defined as the squared norm of the topological term $F\tilde{F}$.
This term has no effect on the purely elelctric or magnetic RN black hole, but it could modify  the dyonic  RN solution completely.
As will be shown in Eq. (\ref{f-series}), it breaks the electromagnetic duality and the electric and magnetic charges $(Q,P)$ enter the standard dyonic black hole asymmetrically.
In this sense, the coupling  of a scalar to  the quasi-topological term is more interesting than the topological term because the latter does not  modify a solution.

We  derive  the Einstein  equation from the action (\ref{Act1})
\begin{eqnarray}
 R_{\mu\nu}-\frac{R}{2}g_{\mu\nu}=2\partial _\mu \phi\partial _\nu \phi -(\partial \phi)^2g_{\mu\nu}+2T_{\mu\nu}, \label{equa1}
\end{eqnarray}
where the energy-momentum tensor takes the form
\begin{eqnarray}
T_{\mu\nu}&=&e^{\alpha \phi^2}\Big(F_{\mu\rho}F_{\nu}~^\rho-\frac{F^2}{4}g_{\mu\nu}\Big) \nonumber \\
          &+&2F^2F_{\mu\rho}F^\rho_\nu-4F_{\mu\rho} F^\rho_\sigma F^\sigma_\lambda F^\lambda_\nu-\frac{1}{4}\Big((F^2)^2-2F^{(4)}\Big)g_{\mu\nu} \label{emten}.
\end{eqnarray}
The Maxwell equation leads to
\begin{eqnarray} \label{M-eq}
&&\nabla_\mu F^{\mu\nu}+2\alpha \phi\nabla_{\mu} (\phi)F^{\mu\nu}+2\nabla_\mu(F^2F^{\mu\nu}-2F^{\mu\rho} F^\sigma_\rho F_{\sigma}^{\nu})=0.\label{M-eq1}
\end{eqnarray}
Importantly, the scalar equation is given by
\begin{equation}
\square \phi -\frac{\alpha}{2}F^2e^{\alpha \phi^2}\phi=0 \label{s-equa}.
\end{equation}
First of all, a dyonic RN  black hole solution without scalar hair could be found as
\begin{eqnarray}
&&ds^2_{\rm dRN}=\bar{g}_{\mu\nu}dx^\mu dx^\nu=-f(r) dt^2+\frac{dr^2}{f(r)} +r^2d\Omega^2_2,  \nonumber \\
&& f(r)=1-\frac{2M}{r}+\frac{P^2}{r^2}+\frac{Q^2}{r^2} {}_2F_1\Big[\frac{1}{4},1,\frac{5}{4};-\frac{4P^2}{r^4}\Big], \label{dRN-bh}\\
&& \bar{\phi}(r)=0,\quad \bar{A}=\bar{v}(r)dt+P\cos\theta d\varphi. \nonumber
\end{eqnarray}
which  is obtained, irrespective of  any value of $\alpha$.
For large $r$, the metric function behaves as
 \begin{equation} \label{f-series}
  f(r)=1-\frac{2M}{r}+\frac{Q^2+P^2}{r^2}-\frac{4Q^2P^2}{5r^6}+\frac{16Q^2P^4}{9r^{10}}-\frac{64 Q^2P^6}{13 r^{14}}+\cdots,
 \end{equation}
 where the first-three terms represent the metric for the standard dyonic RN black hole.
This black hole could be described by a set of parameter $(M,Q,P)$
where $M$  denotes the ADM mass, $Q$ is the electric charge, and $P$ is the magnetic charge.  Two horizons $r=r_\pm$ will be determined by imposing $f(r)=0$.
In addition, we have  $\bar{\phi}=0$ and $\bar{A}_\mu dx^\mu =\bar{v}(r)dt+ P \cos \theta d\varphi$ with $\bar{v}(r)= {}_2F_1[1/4,1,5/4;-4P^2/r^4]Q/r-\bar{\Phi}_Q$.
Here we choose $\bar{\Phi}_Q={}_2F_1[1/4,1,5/4;-4P^2/r_+^4]Q/r_+$ to impose the condition of $\bar{v}(r_+)=0$.
It is worth mentioning  that the non-zero constant $P\not=0$  together with quasi-topological term determines  the solution (\ref{dRN-bh}) uniquely.
In other words, for $P=0$,  the  quasi-topological term gives neither contribution to the energy-momentum tensor (\ref{emten}) nor to the Maxwell equation (\ref{M-eq1}).
We note again that this term has no dynamical effect on the solution of the RN black hole with electric charge.
Hereafter,  we are interested in the dyonic RN black holes with two horizons where could be obtained by choosing ($M,Q,P$) appropriately.
Further, we consider only the region on and outside the outer horizon ($r\ge r_+$).

Introducing  perturbations around the background configuration
\begin{equation}
g_{\mu\nu}=\bar{g}_{\mu\nu}+h_{\mu\nu},\quad\phi=0+\delta \varphi, \quad F_{\mu\nu}=\bar{F}_{\mu\nu}+f_{\mu\nu},\quad f_{\mu\nu} =\partial_\mu a_\nu-\partial_\nu a_\mu,
\end{equation}
we derive the linearized scalar equation as
\begin{equation}
\Big[\bar{\square}+ \alpha \Big\{\frac{Q^2 r^4}{(4P^2+r^4)^2}-\frac{P^2}{r^4}\Big\}\Big]\delta \varphi=0.\label{per-eq}
\end{equation}
In  limit of $P\to 0$, one recovers the scalar equation found in the EMS theory with electric charge.
Now, we focus on the the linearized scalar equation (\ref{per-eq}) which determines totally the instability of dyonic RN black hole  found from the ESM theory with quasi-topological term.
\begin{figure*}[t!]
   \centering
  \includegraphics{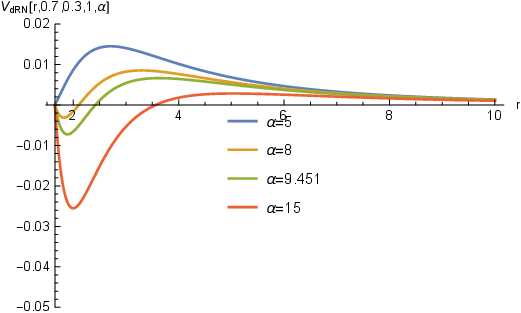}
\caption{The $\alpha$-dependent potentials as function of $r\in [r_+,10]$  with  the outer horizon radius $r_+=1.652(Q=0.7,P=0.3)$ and $l=0$.
 Curves represent the potential  $V_{\rm dRN}(r,0.7,0.3,1,\alpha)$ with different $\alpha=5,8,\alpha_{\rm th}=9.451,15$. }
\end{figure*}
Considering separation of variables
 \begin{equation}
 \delta \varphi(t,r,\theta,\phi) =\int \sum_{lm} \varphi(r) Y_{lm}(\theta) e^{i m \phi} e^{-i\omega t} d\omega,\quad \varphi(r)=\frac{u(r)}{r},
\end{equation}
Eq. (\ref{per-eq}) takes the Schr\"{o}dinger-equation with the tortoise coordinate $r_*$
\begin{equation}
\frac{d^2 u(r)}{dr^2_*} +\Big[ \omega^2-V_{\rm dRN}(r,Q,P,M,\alpha)\Big] u(r) =0, \quad r_*=\int \frac{dr}{f(r)}.
\end{equation}
Here,  the potential is given by
\begin{eqnarray}
V_{\rm dRN}(r,Q,P,M,\alpha)&=&f(r)\Big[-\frac{1}{4P^2r^4+r^8}\Big(8P^2+r^4(Q^2-2Mr)+2P^2r(-4M+r^3)\\ \nonumber
&+&Q^2(4P^2+r^4)~{}_2F_1[\frac{1}{4},1,\frac{5}{4};-\frac{4P^2}{r^4}]\Big)+\frac{l(l+1)}{r^2}-\frac{\alpha Q^2 r^4}{(4P^2+r^4)^2}+\frac{\alpha P^2}{r^4}\Big],\label{RN-P}
\end{eqnarray}
where the case of $Q^2>P^2$ induces the tachyonic instability depending on the coupling parameter $\alpha$. In limit of $P\to 0$, one recovers the scalar potential for the EMS theory. We confine ourselves to the case of  $Q^2>P^2$ because the other case of  $Q^2<P^2$  may induce a positive definite potential, leading to the stable dyonic RN  black hole. Hereafter, we introduce an $s(l=0)$-mode  scalar perturbation for a further consideration.

Fig. 1 suggests that  the threshold of instability $\alpha_{\rm th}$ is  between $\alpha=8$ and  $\alpha=15$ for $Q=0.7$ and $P=0.3$.
To determine the threshold of instability $\alpha_{\rm th}$ precisely, one has to solve the second-order differential equation numerically
\begin{equation}\label{pertur-eq}
\frac{d^2u}{dr_*^2}-\Big[\Omega^2+V_{\rm dRN}(r)\Big]u(r)=0,
\end{equation}
which  allows an exponentially growing mode of  $e^{\Omega t}(\omega_i=\Omega>0) $ as  an unstable mode for $\omega=\omega_r+i\omega_i$ with $\omega_r=0$.
Here we choose two boundary conditions: a normalizable
solution of $u(\infty)\sim e^{-\Omega r_*}$  at infinity  and
a solution of $u(r_+)\sim \left(r-r_+\right)^{\Omega r_+}$  near the outer horizon.
\begin{figure*}[t!]
   \centering
  \includegraphics{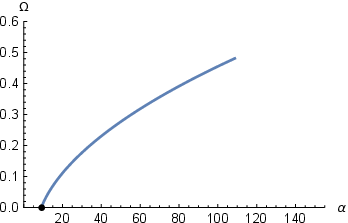}
  \hfill%
\includegraphics{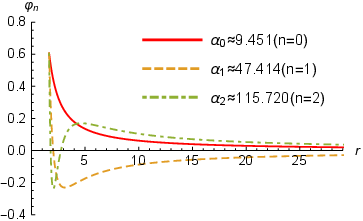}
\caption{
(Left) Plot of an unstable scalar mode with $Q=0.7$ and $P=0.3$. The $y(x)$-axis denote $\Omega$ in $e^{\Omega t}$ $(\alpha)$.
We observe that the threshold of instability $(\bullet:\Omega=0)$ is located at $\alpha_{\rm th}=9.451$.
 (Right) Radial profiles of $\varphi(r)=u(r)/r$ as function of $r\in[r_+=1.651,30]$ for the first three  scalar clouds with $Q=0.7$ and $P=0.3$.
These solutions  $\varphi_n(r)$ are classified by the order number $n=0,1,2$ which is identified by the number of nodes (zero crossings).  }
\end{figure*}
We observe from (Left) Fig. 2 that  the threshold ($\Omega=0$) of instability is located at  $\alpha_{\rm th}=9.451$ for $Q=0.7$ and $P=0.3$.
This implies that  the dyonic RN black hole is unstable for $\alpha>\alpha_{\rm th}$, while it is stable for  $\alpha<\alpha_{\rm th}$.

The other way of obtaining $\alpha_{\rm th}$ is to solve the  static linearized  equation directly.
We consider the  static scalar linearized equation around the dyonic RN black hole background to find the onset of scalarization for $n=0,~ 1,~ 2,\cdots$ black holes as
\begin{equation} \label{ssclar-eq}
\frac{1}{r^2}\frac{d}{dr}\Big[r^2f(r)\frac{d\varphi(r)}{dr}\Big]-\Big[\frac{l(l+1)}{r^2}-\frac{\alpha Q^2 r^4}{(4P^2+r^4)^2}+\frac{\alpha P^2}{r^4}\Big] \varphi(r)=0
\end{equation}
which describes an eigenvalue problem: for a given $l=0$, requiring an asymptotically vanishing, smooth scalar field
selects a discrete set of $n=0$, 1, 2, $\cdots$. Also, these determine the bifurcation points (discrete resonant spectrum: $\{\alpha_n\}$) numerically.
We plot $\varphi_n(r)$ as a function of $r$ with three $n=0(\alpha_0=9.451),~n=1(\alpha_1=47.414),~n=2(\alpha_2=115.72)$ whose forms can be found from (Right) Fig. 2.
We confirm that  $\alpha_{\rm th}=\alpha_0$.
It is interesting to note    that the  scalar cloud $\varphi_0$ without zero crossing will develop the fundamental branch of scalarized charged black hole with $\alpha\ge \alpha_0=\alpha_{\rm th}$,
while the scalar clouds  $\varphi_1,~\varphi_2$ with zero crossings will develop the $n=1,~2$ excited branches of scalarized charged black holes with $\alpha\ge \alpha_1,~\alpha_2$, respectively.   In general, the infinite  $n=0,~1,~2,\cdots$ black holes with $Q=0.7$ and $P=0.3$ are defined by $\alpha$-bounds of $\alpha\ge \alpha_0,~\alpha\ge \alpha_1$, $\alpha\ge \alpha_2,~\cdots$, respectively.

\subsection{Scalarized charged black holes}
 All scalarized charged black holes will  be generated from the onset of scalarization $\{\varphi_n(r)\}$ in the unstable region of dyonic RN black hole ($\alpha\ge \alpha_{\rm th}$). It is clear that the scalarized solutions arise dynamically from the evolution of dyonic RN black hole when a scalar cloud (perturbation) plays the role of a seed.

Let us  obtain scalarized charged black holes through spontaneous scalarization.
For this purpose,  we consider  the metric and fields as~\cite{Herdeiro:2018wub}
\begin{eqnarray}\label{nansatz}
ds^2_{\rm SCBH}&=&-N(r)e^{-2\delta(r)}dt^2+\frac{dr^2}{N(r)}+r^2(d\theta^2+\sin^2\theta d\varphi^2) \nonumber \\
N(r)&=&1-\frac{2m(r)}{r},\quad \phi=\phi(r),\quad A=v(r)dt+P\cos\theta d\varphi.
\end{eqnarray}
Plugging (\ref{nansatz}) into (\ref{equa1})-(\ref{s-equa}), one finds the four equations
\begin{eqnarray}
&&e^{\alpha\phi(r)^2}P^2-2r^2m'(r)+e^{2\delta(r)}(4P^2+e^{\alpha\phi(r)^2}r^4)(v'(r))^2\nonumber\\
&&+r^3\left(r-2m(r)\right)\phi'(r)^2=0, \label{neom1}\\
&&\delta'(r)+r(\phi'(r))^2=0, \label{neom2}\\
&&v'(r)+\frac{Qr^2}{e^{\delta(r)}(4P^2+e^{\alpha\phi(r)^2}r^4)}=0, \label{neom3}\\
&&-e^{\alpha\phi(r)^2}\alpha\phi(r)[P^2-e^{2\delta(r)}r^4(v'(r))^2]-2r^2[m(r)+rm'(r)-r]\phi'(r)\nonumber\\
&&-r^3(r-2m(r))[\delta'(r)\phi'(r)-\phi''(r)]=0, \label{neom4}
\end{eqnarray}
where the prime ($'$) denotes differentiation with respect to $r$.

Implementing the existence of a horizon located at $r=r_+$ with $v(r_+)=0$,   an
approximate solution to equations (\ref{neom1})-(\ref{neom4}) is found  in the near-horizon as
\begin{eqnarray}
m(r)&=&\frac{r_+}{2}+m_1(r-r_+)+\cdots,\quad
\delta(r)=\delta_0+\delta_1(r-r_+)+\cdots,\label{aps-1}\\
\phi(r)&=&\phi_0+\phi_1(r-r_+)+\cdots,\quad v(r)=v_{1}(r-r_+)+\cdots,\label{aps-2}
\end{eqnarray}
where the four coefficients are given by
\begin{eqnarray}\label{ncoef}
&&m_1=\frac{e^{\alpha\phi_0^2}P^2(4P^2+e^{\alpha\phi_0^2} r_+^4)+Q^2r_+^4}{2r_+^2(4P^2+e^{\alpha\phi_0^2}r_+^4)},\quad
\delta_1=-r_+\phi_1^2,\nonumber\\
&&\phi_1=\frac{e^{\alpha\phi_0^2}[Q^2r_+^8-P^2(4P^2+e^{\alpha\phi_0^2}r_+^4)^2]\alpha\phi_0}{r_+(4P^2
+e^{\alpha\phi_0^2}r_+^4)[(4P^2
+e^{\alpha\phi_0^2}r_+^4)(e^{\alpha\phi_0^2}P^2-r^2_+)+Q^2r_+^4]},\nonumber\\
&&v_1=-\frac{e^{-\delta_0}Qr_+^2}{(4P^2+e^{\alpha\phi_0^2}r_+^4)}.
\end{eqnarray}
Here, we note that ``$4P^2$" in $(4P^2+e^{\alpha\phi_0^2}r_+^4)$ is a newly contribution to four coefficients when comparing with the EMS theory.
Also, two important parameters of $\phi_0=\phi(r_+,\alpha)$ and $\delta_0=\delta(r_+,\alpha)$ are
determined when matching with an asymptotically flat solution in the far-region
\begin{eqnarray}\label{ncoef}
&&m(r)=M-\frac{P^2+Q^2+Q_s^2}{2r}+\cdots,\quad
\delta(r)=\frac{Q_s^2}{2r^2}+\cdots,\nonumber\\
&&v(r)=-\Phi_Q+\frac{Q}{r}+\cdots,\quad
\phi(r)=\frac{Q_s}{r}+\cdots,
\end{eqnarray}
where  $Q_s$ and  $\Phi_Q$ represent the scalar charge and the electrostatic potentials at infinity,
in addition to the ADM mass $M$, and the electric charge $Q$ and magnetic charge $P$.

\begin{figure*}[t!]
   \centering
  \includegraphics{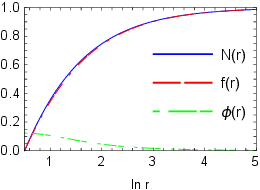}
  \hfill%
  \includegraphics{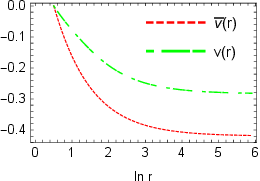}
    \hfill%
  \includegraphics{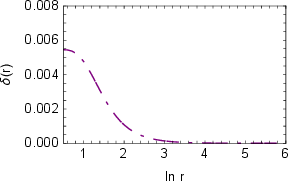}
\caption{Plots of a scalarized charged black hole with $\alpha=54$ belonging to  the $n=0$  branch of   $\alpha \ge \alpha_0=9.451$.
 Here,  the horizon is located  at $r=r_+=1.651$ ($\ln r_+=0.503$) and two parameters are given by   $\phi_0=0.1231$ and $\delta_0=0.0054$.
 $f(r)$ and $\bar{v}(r)$ represent those for  the dyonic RN black with $\bar{\phi}(r)=\bar{\delta}(r)=0$. }
\end{figure*}
Regarding as  an explicit scalarized charged black hole solution with $P=0.3$ and $Q=0.7$,
we present a numerical black hole solution with $\alpha=54$, and mass $M=0.9812$ in the $n=0$ fundamental  branch of $\alpha\ge 9.451$ in Fig 3.
Further, one needs  to explore  hundreds of numerical solutions depending $\alpha$ for each branch to perform the stability of scalarized charged black holes.

\subsection{Stability of scalarized charged black holes}

Before we proceed, it is noted that the stability analysis for scalarized charged black holes is an important task
since it determines their viability in representing realistic astrophysical configurations.
The conclusions about the stability of the scalarized charged black holes with respect to  perturbations will be reached by examining
the qualitative behavior of the potential  as well as by obtaining exponentially growing (unstable) modes for $s$-mode scalar  perturbation.

Here, we prefer to  introduce the radial perturbations around the scalarized black holes as
\begin{eqnarray}
&&ds_{\rm rad}^2=-N(r)e^{-2\delta(r)}(1+\epsilon H_0)dt^2+\frac{dr^2}{N(r)(1+\epsilon H_1)}
+r^2(d\theta^2+\sin^2\theta d\psi^2),\nonumber\\
&&F_{tr}(t,r)=v'_{Q}(r)+\epsilon\delta v_Q(t,r),\quad
\phi(t,r)=\phi(r)+\epsilon\delta\tilde{\phi}(t,r), \label{p-metric}
\end{eqnarray}
where $N(r)$, $\delta(r)$, $v_{Q}(r)$, and  $\phi(r)$ represent a scalarized
charged black hole background, while
$H_0(t,r)$, $H_1(t,r)$, $\delta v_{Q}(t,r)$, and $\delta\tilde{\phi}(t,r)$
denote four perturbed fields around the scalarized
black hole background. From now on, we focus on  the $l=0$(s-mode)
scalar propagation  by mentioning  that higher angular momentum modes $(l\neq0)$ are neglected. In this
case, other three perturbed fields  become redundant fields.

Seeking for a decoupling process by using  linearized equations, one may find a linearized scalar equation.
Considering the separation of variables
\begin{eqnarray}
\delta\tilde{\phi}(t,r)=\frac{\tilde{\varphi}(r)e^{\Omega t}}{r},
\end{eqnarray}
we obtain the Schr\"{o}dinger-type equation for an $s$-mode scalar perturbation
\begin{eqnarray}
\frac{d^2\tilde{\varphi}(r)}{dr_*^2}-\Big[\Omega^2+V_{\rm mt}(r,\alpha)\Big]\tilde{\varphi}(r)=0,
\end{eqnarray}
with $r_*$ is the tortoise coordinate defined by
\begin{eqnarray}
\frac{dr_*}{dr}=\frac{e^{\delta(r)}}{N(r)}.
\end{eqnarray}
Here, its potential reads to be
\begin{eqnarray} \label{sc-poten}
V_{\rm mt}(r,\alpha)&=&\frac{Ne^{-2\delta}}{r^2}\Big[(1-N-2r^2\phi'^2)
+\frac{e^{\alpha\phi^2}P^2[\alpha-1+2(\alpha\phi+r\phi')^2]}{r^2}\Big]\nonumber\\
&+&\frac{Ne^{-2\delta}Q^2}{\left(4P^2+e^{\alpha\phi^2}r^4\right)^3}\left(8P^2(2P^2+e^{\alpha\phi^2}r^4)(2r^2\phi'^2-1)
-4\alpha e^{\alpha\phi^2}P^2r^4(1+2\alpha\phi^2+4r\phi\phi')\right.\nonumber\\
&&\left.+e^{2\alpha\phi^2}r^8(-\alpha-1+2(-\alpha\phi+r\phi')^2])\right),
\end{eqnarray}
whose limit of $P^2\to 0$ recovers the potential for the EMS theory.
\begin{figure*}[t!]
\centering
\subfigure[]{
\includegraphics{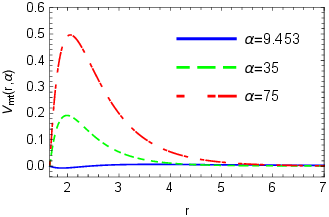}}
\hfill%
\subfigure[]{
\includegraphics{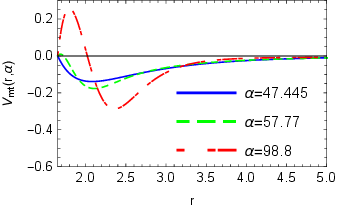}}
\hfill%
\subfigure[]{
\includegraphics{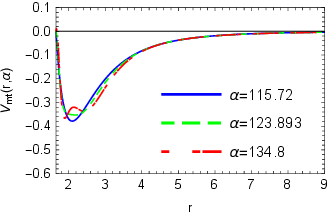}}
\caption{ Scalar potentials $V_{\rm mt}(r,\alpha)$ around $n=0$ [(a): $\alpha\ge9.451$], 1 [(b): $\alpha\ge47.414$],
2 [(c): $\alpha\ge115.72$] black holes in the infinite branches. The positive barriers in the near-horizon become smaller and smaller as $n$ increases.  }
\end{figure*}
At this stage, we wish to analyze the potential $V_{\rm mt}(r,\alpha)$ carefully.
First of all, we observe  that the second line from quasi-topological term  contributes to a negatively small potential in the near-horizon, while the first and last lines are combined to give  the potential for the EMS theory based on the standard dyonic RN black hole.
In Fig. 4, as the number ($n$) of scalar-node increases, the negative region of $V(r,\alpha)$ increases in the near-horizon.
We display three scalar potentials $V_{\rm mt}(r,\alpha)$ in  Fig. 4(a) for $l=0(s$-mode) scalar around the $n=0$ black hole, showing positive definite.
This implies that the $n=0$ black hole is  stable against the $s$-mode of perturbed scalar.
 We observe from  Fig. 4(b) and 4(c) that $\int_{r_+}^\infty dr[e^\delta V(r,\alpha)/N]<0$ (sufficient condition for instability~\cite{Dotti:2004sh}) for the $n=1,~2$ black  holes.
It suggests that the $n=1,~2$ black holes may be  unstable against the $s$-mode scalar perturbation.
\begin{figure*}[t!]
   \centering
  \includegraphics{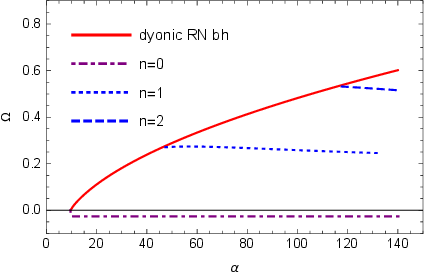}
  \hfill%
  \includegraphics{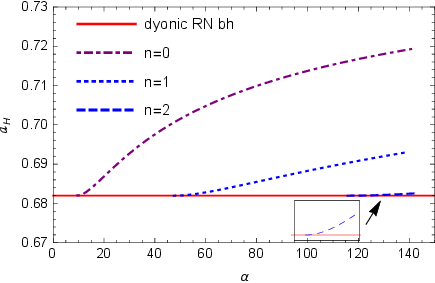}
\caption{(Left) Plots of $\omega_i=\Omega(\omega_r=0)$ as functions of $\alpha$ for $l=0$-scalar mode around the $n=0(\alpha\ge9.451)$,
$1(\alpha\ge47.414)$, $2(\alpha\ge115.72)$ black holes with $Q=0.7$ and $P=0.3$. (Right) Plots of the reduced entropy $a_{\rm H}$  as a function of $\alpha$ with fixed  $Q=0.7$
and $P=0.3$.}
\end{figure*}
Actually, we confirm from (Left) Fig. 5 that  the positive $\Omega$ for $n=1$, 2 black holes implies unstable
black holes, while the negative $\Omega$ for the $n=0$ black hole shows a stable black hole. A red curve starting at $\alpha=9.451(\Omega=0)$
denotes the positive $\Omega$, showing the unstable dyonic RN black holes for $\alpha>9.451$.
All $n=0,~1,~2,~\cdots$ scalarized black holes are started from bifurcation points on the red curve.

Finally, let us discuss  the entropic preference for $n=0,1,2$ scalarized black holes and dyonic RN black hole~\cite{Astefanesei:2019pfq}.
For this purpose, it is convenient to use the  reduced entropy defined by
\begin{equation} \label{entropy}
a_{\rm H}=\frac{A_{\rm H}}{16\pi M^2}
\end{equation}
 with $A_{\rm H}=4\pi r_+^2$.
As is shown (Right) Fig.5,  for $Q=0.7$ and $P=0.3$,  the $n=0,1,2$ scalarised black holes are always entropically preferred than the dyonic RN black hole.
 In addition, we  observe that $a_{\rm H}$ increases with the growth of $\alpha$.
 However, we wish to point out  that the entropic preferences for $n=1,2$ scalarized black holes have nothing to do with their dynamic instabilities.

\section{Scalar coupling to quasi-topological term}

In this section, we perform briefly the whole process of section 2 for the scalar coupling to quasi-topological electromagnetism given by the action
\begin{equation}
S_{\rm EMSqT}=\frac{1}{16 \pi}\int d^4 x\sqrt{-g}\Big[ R-2\partial_\mu \phi \partial^\mu \phi- F^2-e^{\alpha \phi^2}\Big((F^2)^2-2F^{(4)}\Big)\Big].\label{Act2}
\end{equation}
Relevantly, a scalar equation takes the form
\begin{equation}
\square \phi -\frac{\alpha}{2}\Big((F^2)^2-2F^{(4)}\Big)e^{\alpha \phi^2}\phi=0 \label{st-equa}.
\end{equation}

We need the perturbed scalar equation to study the tachyonic instability of the dyonic RN black hole as
\begin{equation}
\Big[\bar{\square}+ \alpha\frac{4Q^2P^2 }{(4P^2+r^4)^2}\Big]\delta \varphi=0,\label{tper-eq}
\end{equation}
which implies that the tachyonic instability always appears for any $Q$ and $P$.
Before we proceed, if one includes a scalar coupling to the topological term like  $e^{\alpha \phi^2}F\tilde{F}$ instead of the scalar coupling to the quasi-topological term,
its linearized equation is given by
\begin{equation}
\Big[\bar{\square}+ \alpha\frac{2QP }{4P^2+r^4}\Big]\delta \varphi=0,\label{toper-eq}
\end{equation}
where the coupling term is the square-root of the quasi-topological term. We note  that (\ref{toper-eq}) is similar to (\ref{tper-eq}).

To find the bifurcation points numerically, we need to have the static perturbed  equation based on the dyonic RN black hole as
\begin{equation} \label{tssclar-eq}
\frac{1}{r^2}\frac{d}{dr}\Big[r^2f(r)\frac{d\varphi(r)}{dr}\Big]-\Big[\frac{l(l+1)}{r^2}-\frac{4\alpha Q^2 P^2}{(4P^2+r^4)^2}\Big] \varphi(r)=0.
\end{equation}
Here, we find the first three bifurcation points as $\alpha_0=4.933,~\alpha_1=18.687,~\alpha_2=45.590$
with $Q=0.7$ and $P=0.8$ (see Fig. 6).
It is curious to note    that the $\varphi_0$ scalar cloud  will develop the fundamental branch of scalarized charged black hole with $\alpha\ge \alpha_0=\alpha_{\rm th}$,
while the $\varphi_1,~\varphi_2$ scalar clouds  will develop the $n=1,~2$ excited branches of scalarized charged black holes with $\alpha\ge \alpha_1,~\alpha_2$, respectively.
We note that all scalarized black holes would be generated from the onset of scalarization $\{\varphi_n(r)\}$  in the unstable region ($\alpha\ge \alpha_{\rm th}$) of dyonic RN black hole.
\begin{figure*}[t!]
   \centering
\includegraphics{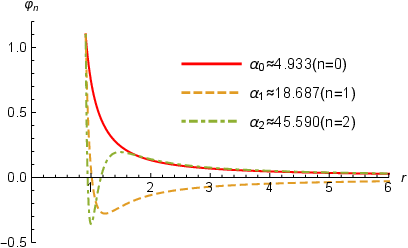}
\caption{
 Radial profiles of $\varphi_n(r)$ as function of $r\in[r_+=0.9078,6]$ for the first three perturbed scalar solutions with $Q=0.7$ and $P=0.8$. }
\end{figure*}

We  develop infinite scalarized charged black holes through spontaneous scalarization followed by section 2.2.
As  an explicit scalarized charged black hole solution with $Q=0.7$, and $P=0.8$,
we find a numerical black hole solution
with $\alpha=54$, and mass $M=0.9512$ in the $n=0$ fundamental  branch of $\alpha\ge \alpha_0$.
Here we do not wish to  display it because it takes a similar form as in Fig. 3.
However, we have  to generate  hundreds of numerical solutions depending $\alpha$ for each branch to perform the stability of these scalarized charged black holes.

Making use of a decoupling process on linearized equations, one may find a linearized scalar equation.
Introducing
\begin{eqnarray}
\delta\tilde{\phi}(t,r)=\frac{\tilde{\varphi}(r)e^{\Omega t}}{r},
\end{eqnarray}
we obtain the Schr\"{o}dinger-type equation for an $s$-mode scalar perturbation
\begin{eqnarray}
\frac{d^2\tilde{\varphi}(r)}{dr_*^2}-\Big[\Omega^2+V_{\rm qtt}(r,\alpha)\Big]\tilde{\varphi}(r)=0.
\end{eqnarray}
Here, its potential reads to be
\begin{eqnarray} \label{sc-poten}
V_{\rm qtt}(r,\alpha)&=&Ne^{-2\delta}\Big[-\frac{N}{r^2}+\frac{(P^2-r^2)(2r^2\phi'^2-1)}{r^4}
+\frac{Q^2\left(2(\alpha\phi-r\phi')^2-\alpha-1\right)}{4e^{\alpha\phi^2}P^2+r^4}\nonumber\\
&+&\frac{Q^2r^4\alpha(1-6\alpha\phi^2+4r\phi\phi')}{\left(4e^{\alpha\phi^2}P^2+r^4\right)^2}
+\frac{4\alpha^2Q^2r^8\phi^2}{\left(4e^{\alpha\phi^2}P^2+r^4\right)^3}\Big].
\end{eqnarray}

\begin{figure*}[t!]
\centering
\subfigure[]{
\includegraphics{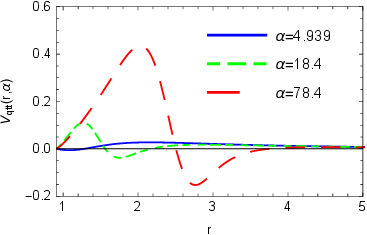}}
\hfill%
\subfigure[]{
\includegraphics{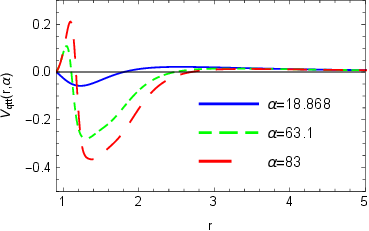}}
\hfill%
\subfigure[]{
\includegraphics{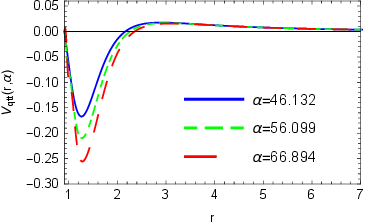}}
\caption{Scalar potentials $V_{\rm qtt}(r,\alpha)$ around $n=0$ [(a): $\alpha\ge4.933$], 1 [(b): $\alpha\ge18.687$],
2 [(c): $\alpha\ge45.590$] black holes. The positive barriers in
the near-horizon become smaller and smaller as $n$ increases.  }
\end{figure*}
As is shown in Fig. 7, the potentials $V_{\rm qtt}$ for $n=1,2$  black holes are similar to $V_{\rm mt}$ for $n=1,2$ black holes in Fig. 4.
The potential $V_{\rm qtt}$ for $n=0$ black hole is slightly different from $V_{\rm mt}$ for $n=0$ black hole, implying that it may be stable against the $s$-mode scalar perturbation.
We confirm from (Left) Fig. 8 that  the positive $\Omega$ for $n=1$, 2 black holes implies unstable
black holes, while the negative $\Omega$ for the $n=0$ black hole shows a stable black hole.
\begin{figure*}[t!]
   \centering
  \includegraphics{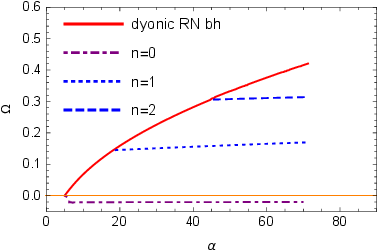}
  \hfill%
  \includegraphics{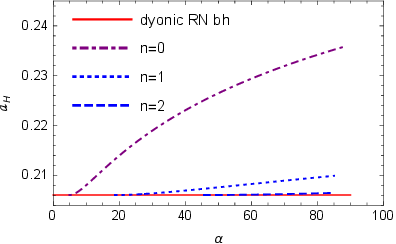}
\caption{(Left) Plots of $\omega_i=\Omega$ as functions of $\alpha$ for $l=0$-scalar mode around the $n=0(\alpha\ge4.933)$,
$1(\alpha\ge18.687)$, $2(\alpha\ge45.590)$ black holes with $Q=0.7$ and $P=0.8$. (Right) Plots of the reduced entropy $a_{\rm H}$  as a function of $\alpha$ with fixed  $Q=0.7$
and $P=0.8$. }
\end{figure*}

Lastly, we comment on  the entropic preference for $n=0,1,2$ scalarized black holes and dyonic RN black hole.
As is shown (Right) Fig. 8,  for $Q=0.7$ and $P=0.8$,  the $n=0,1,2$ scalarised black holes are  entropically favored than the dyonic RN black hole.
It is curious  to mention that the entropic preferences for $n=1,2$ scalarized black holes have nothing to do with their dynamic instabilities.

\section{Scalar coupling to Maxwell and  quasi-topological terms}
Here, we mention briefly the process of section 2 for the scalar coupling to Maxwell and quasi-topological terms given by the action
\begin{equation}
S_{\rm EMSQT}=\frac{1}{16 \pi}\int d^4 x\sqrt{-g}\Big[ R-2\partial_\mu \phi \partial^\mu \phi- e^{\alpha \phi^2}F^2-e^{\alpha \phi^2}\Big((F^2)^2-2F^{(4)}\Big)\Big].\label{Act3}
\end{equation}
A scalar equation takes the form
\begin{equation}
\square \phi -\frac{\alpha}{2}F^2e^{\alpha \phi^2}\phi -\frac{\alpha}{2}\Big((F^2)^2-2F^{(4)}\Big)e^{\alpha \phi^2}\phi=0 \label{bst-equa}.
\end{equation}
A dyonic RN  black hole solution to (\ref{Act3}) without scalar hair is also given by (\ref{dRN-bh}).
One  needs a perturbed scalar equation to study the tachyonic instability of the dyonic RN black hole as
\begin{equation}
\Big[\bar{\square}+ \alpha\Big(\frac{Q^2 }{4P^2+r^4}-\frac{P^2}{r^4}\Big)\Big]\delta \varphi=0,\label{ttper-eq}
\end{equation}
which means that the tachyonic instability may be allowed  for  $Q>P$.
To find the bifurcation points numerically, we need to solve the static perturbed  equation based on the dyonic RN black hole as
\begin{equation} \label{tssclar-eq}
\frac{1}{r^2}\frac{d}{dr}\Big[r^2f(r)\frac{d\varphi(r)}{dr}\Big]-\Big[\frac{l(l+1)}{r^2}- \frac{\alpha Q^2 }{4P^2+r^4}+\frac{\alpha P^2}{r^4}\Big] \varphi(r)=0.
\end{equation}
We find the first three bifurcation points as $\alpha=\alpha_0=9.187,~\alpha_1=46.236,~\alpha_2=112.901$
with $Q=0.7$ and $P=0.3$. Also, we obtain three scalar clouds $\varphi_n$  for $n=0,1,2$ in Fig. 9.
\begin{figure*}[t!]
   \centering
  \includegraphics{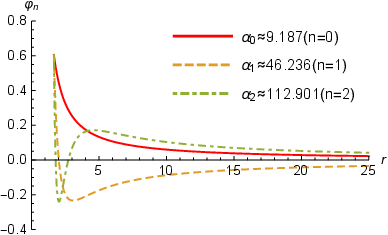}
\caption{Radial profiles of $\varphi_n(r)$ as function of $r\in[r_+=?,6]$ for the first three perturbed scalar solutions with $Q=0.7$ and $P=0.3$. }
\end{figure*}
We  make  infinite scalarized charged black holes through spontaneous scalarization followed by section 2.2.
As  an explicit scalarized charged black hole solution with $Q=0.7$ and $P=0.3$,
we  may find a numerical black hole solution with
$\alpha=54$, and mass $M=0.9808$ in the $n=0$ fundamental  branch of $\alpha\ge \alpha_0$.
Here we do not want  to  display it because it takes a similar form as in Fig. 3.
However, we should generate  hundreds of numerical solutions depending $\alpha$ for each branch to perform the stability of these scalarized charged black holes.

Considering  a decoupling process on linearized equations, one may find a linearized scalar equation.
Introducing
\begin{eqnarray}
\delta\tilde{\phi}(t,r)=\frac{\tilde{\varphi}(r)e^{\Omega t}}{r},
\end{eqnarray}
we obtain the Schr\"{o}dinger-type equation for an $s$-mode scalar perturbation
\begin{eqnarray}
\frac{d^2\tilde{\varphi}(r)}{dr_*^2}-\Big[\Omega^2+V_{\rm mqt}(r,\alpha)\Big]\tilde{\varphi}(r)=0.
\end{eqnarray}
Here, its potential reads to be
\begin{eqnarray}
V_{\rm mqt}(r,\alpha)&=&\frac{Ne^{-2\delta}}{r^2}\Big[1-N-2r^2\phi'^2
+\frac{e^{\alpha\phi^2}P^2[\alpha-1+2(\alpha\phi+r\phi')^2]}{r^2}\nonumber\\
&+&\frac{e^{-\alpha\phi^2}Q^2r^2[-\alpha-1+2(-\alpha\phi+r\phi')^2]}{4P^2+r^4}\Big].
\end{eqnarray}
As is displayed in Fig. 10, the potentials $V_{\rm mqt}$ are quite  similar to  $V_{\rm mt}$ in Fig. 4.
This implies that  $V_{\rm mqt}$ is determined significantly  by the Maxwell term rather than the quasi-topological term.

\begin{figure*}[t!]
\centering
\subfigure[]{
\includegraphics{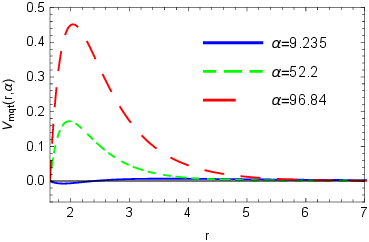}}
\hfill%
\subfigure[]{
\includegraphics{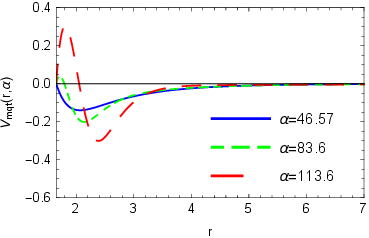}}
\hfill%
\subfigure[]{
\includegraphics{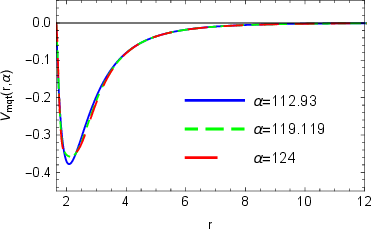}}
\caption{ Scalar potentials $V_{\rm mqt}(r,\alpha)$ around $n=0$ [(a): $\alpha\ge9.187$], 1 [(b): $\alpha\ge46.236$],
2 [(c): $\alpha\ge112.901$] black holes. The positive barriers in
the near-horizon become smaller and smaller as $n$ increases.}
\end{figure*}
From  Fig.10, one may find the $n=0$ black hole is  stable against the $s$-mode of perturbed scalar, while
the $n=1,~2$ black holes are  unstable against the $s$-mode scalar perturbation.
We obtain  from   (Left)Fig. 11 that  the positive $\Omega$ for $n=1$, 2 black holes implies unstable
black holes, while the negative $\Omega$ for the $n=0$ black hole shows a stable black hole.
\begin{figure*}[t!]
   \centering
  \includegraphics{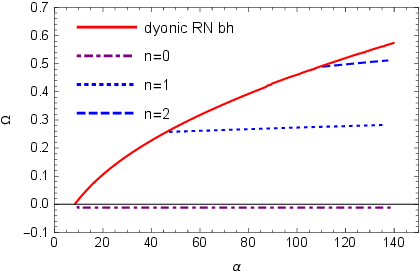}
  \hfill%
  \includegraphics{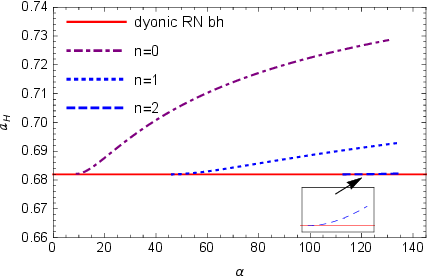}
\caption{ (Left) Plots of $\omega_i=\Omega$ as functions of $\alpha$ for $l=0$-scalar mode around the $n=0(\alpha\ge9.1987)$,
$1(\alpha\ge46.236)$, $2(\alpha\ge112.901)$ black holes with $Q=0.7$ and $P=0.3$. (Right) Plots of the reduced entropy $a_{\rm H}$  as a function of $\alpha$ with fixed  $Q=0.7$
and $P=0.3$. }
\end{figure*}

Finally, we mention the entropic preference for $n=0,1,2$ scalarized black holes and dyonic RN black hole.
As is shown (Right) Fig. 11,  for $Q=0.7$ and $P=0.3$,  the $n=0,1,2$ scalarised black holes are  entropically preferred than the dyonic RN black hole.
We would like to say that the entropic preferences for $n=1,2$ scalarized black holes have nothing to do with their dynamic instabilities.

\section{Discussions }
We  have studied the EMS theory with quasi-topological term.
Introducing exponential coupling $e^{\alpha \phi^2}$ to Maxwell term, quasi-topological term, and both terms,  we obtain three sets of  $n=0,1,2,\cdots$ scalarized charged  black holes
by taking into account tachyonic instability of dyonic RN black hole.
The stability of all scalarized charged black holes with respect to radial perturbations are reached by examining
the qualitative behavior of the potential  as well as by obtaining  exponentially growing  modes for $s$-mode scalar  perturbation.
We find that all $n=0$ black holes are stable, whereas all $n=1,2$ black holes are unstable.
This is consistent with the results for the EMS theory with exponential coupling~\cite{Myung:2018jvi} and quadratic coupling~\cite{Myung:2019oua}, and for the Einstein-Maxwell-conformally coupled scalar theory with exponential and quadratic couplings~\cite{Zou:2020zxq}.
This implies that the $n=0$ scalarized  black hole is meaningful among infinite scalarized black holes.
Also, the $n=0$ scalarized black hole is regarded as the endpoint of the evolution of unstable dyonic RN black hole.
From (Right) Fig. 5, (Right) Fig. 8, and (Right) Fig. 11, it is clear that the $n=0$ scalarized  black hole is entropically preferred than the dyonic RN black hole.

At this stage, we wish to ask  a question of why the $n=0$ black hole is always stable.
All $n=0$ black holes include the threshold of tachyonic stability and their region could be extended as a bound like $\alpha \ge \alpha_0$.
One exceptional case  corresponds to the  $n=0$ black hole found from the EGBS theory with quadratic coupling~\cite{Silva:2017uqg,Blazquez-Salcedo:2018jnn}.
This black hole is unstable because its allowed region is not a bound but a band~\cite{Myung:2019oua}. The band appeared because of the regularity condition.

The stability for a quartic coupling in the EMS theory was recently announced by considering full perturbations~\cite{Blazquez-Salcedo:2020jee}
and critical solutions for scalarized black holes with the same coupling were mentioned in~\cite{Blazquez-Salcedo:2020crd}.

 \vspace{1cm}

{\bf Acknowledgments}

 This work was supported by the National Research Foundation of Korea (NRF) grant funded by the Korea government (MOE)
 (No. NRF-2017R1A2B4002057).
 
\end{document}